\documentclass   [a4paper,12pt]{article}
\usepackage{graphicx}
\begin{document}
\begin{center}
%%{\bf\large Hidden-variable theory - the only non-falsified physical alternative?}\\
{\bf\large Hidden-variable theory versus Copenhagen quantum mechanics}

    Milo\v{s} V. Lokaj\'{\i}\v{c}ek, Institute of Physics, v.v.i.
    
     Academy of Sciences of the Czech Republic, Prague
\end{center}

Abstract

The main assumptions the Copenhagen quantum mechanics has been based on will be summarized and the known (not yet decided) contradiction between Einstein and Bohr will be newly analyzed. The given assumptions have been represented basically by time-dependent Schroedinger equation, to which some further assumptions have been added. Some critical comments have been raised against the given mathematical model structure by Pauli (1933) and by Susskind and Glogover (1964). They may be removed if in principle only the Schroedinger equation is conserved and the additional assumptions are abandoned, as shown recently. It seems to be in contradiction to the numerous declarations that the Copenhagen model has been approved by experimental results.
 However, in the most of these experiments only the agreement with the mere Schroedinger equation has been tested. All mentioned assumptions have been tested practically only in the EPR experiment (measurement of coincidence light transmission through two polarizers) proposed originally by Einstein (1935). Also these experimental results have been interpreted as supporting the Copenhagen alternative, which has not been, however, true. In fact the microscopic world may be described only with the help of the hidden-variable theory that is represented by the Schroedinger equation without mentioned additional assumptions, which has the consequence that the earlier gap between microscopic and macroscopic worlds has been removed. The only difference concerns the existence of discrete states. The possibilities of the human reason of getting to know the nature  will be also shortly discussed in the beginning of this contribution.

\section {Introduction}
One of the main physical theories of the twentieth century was Copenhagen quantum mechanics. It was believed that it provided a true picture of the microworld even if it contained a series of logical paradoxes and the given physical characteristics differed significantly from those of macroscopic world observed also with the help of our senses. It is in principle impossible to imagine the macroworld to consist of microscopic objects fulfilling the properties required by the Copenhagen quantum mechanics. And one should ask if it is possible to remove the corresponding discrepancy.

 Some critical points concerning the Copenhagen quantum mechanics were introduced already in the past. And different unsuccessful attempts were done to remove them. The Copenhagen alternative was, of course, often denoted as supported by different experimental data. However, in the most cases only the assumption of the mere Schroedinger equation \cite{sch} has been tested (without adding further assumptions forming Copenhagen alternative). All corresponding assumptions have been involved practically in the EPR experiment only. The results of this experiment have been interpreted, of course, also as supporting the Copenhagen alternative, which 
has been, however, the consequence of two mistakes, as it will be discussed in the following. 

The aim of the following contribution will be to summarize practically all critical points and to gather all arguments that might contribute to the definite solution of the given problem. All related items will be at least shortly explained and discussed, while all necessary details may be found in corresponding references. It will be shown that it is the hidden-variable theory, i.e., the mere time-dependent Schroedinger equation (practically without any additional assumptions) that should be preferred to Copenhagen quantum mechanics as well as to classical physics.

However, before presenting the main results it is purposeful to start with a short discussion concerning the possibilities of our reason to get individual pieces of scientific knowledge when the corresponding approach of knowing is based in principle on the falsification of different hypothetical statements only. There are always two possibilities that must be recognized: what is the certain truth or what is plausible with the possibility of being falsified later. \\

 \section {Scientific knowledge and human reason}
All human knowledge is based on the observation of world and nature, including human beings. One forms different hypotheses, or statements, with the help of logical induction or of intuition. To test if they are true or wrong all possible consequences must be derived on the basis of logical deduction, which represents the main part of activities of contemporary scientists. If one comes to a logical contradiction or to a contradiction to world observation, the given hypothesis (or a corresponding set of hypotheses) must be denoted as falsified; it must be abandoned or modified, so as the contradiction be removed.

However, even if any falsification has not been found the given hypothesis cannot be denoted as verified, as one never knows what will occur in future (the logical process looking for falsification should continue or new experimental data may be obtained). The certain truth consists in the set of all falsifying statements only; we can know with certainty what is not true.

On the other side we are justified and also duty to denote all non-falsified statements or statement sets as plausible, even if it may happen that some statements from different statement sets are in mutual contradiction. In such a case both the alternatives should be further developed until one of them is found as falsified. However, in the past that was not often respected. The first statement found as non-falsified was denoted usually as verified and any other alternative statements were refused on the principle of falsifiability. See also Refs. \cite{conc1,conc2} where the whole problem has been discussed to a greater detail.        
 
It follows from the preceding that also the logical rules represent a set of basic statements that may undergo the falsification process. Until now the two-value logic must be regarded as fully plausible; the more-value logics having been discussed in relation to the quantum mechanics in the last century have remained without any success. \\

\section {Science and philosophy}
It is possible to say that the natural science was founded by Aristotle (383-322 b.C.) as he was the first one who started with systematical observation of the nature. He also proposed the rules of the two-value logic that has represented the decisive basis of all scientific considerations, and made use fully of the causality principle. 

His considerations were involving always metaphysical problems, which was related to the fact that all physical phenomena started from the ontological structure of matter objects. His approaches (including logical rules) represented the basis of scientific considerations during the whole middle age even if his purely physical statements were falsified at that time. 

In the first millennium the approaches of Aristotle were not practically known in Europe; the preference was given to ideas of Plato. They were brought to Europe through Islam and Arabs in the beginning of the second millennium. Albert the Great (1193-1280) overtook them and Thomas Aquinas (1225-74) continued in developing them further. It is possible to say that all scientific knowledge of the middle age was based on the unity of science and ontological metaphysics, which started to change with the coming of the new age.

Even if it is evident that Thomas Aq. was aware at least in principle that our rational knowledge was based on falsification approach, it has not been explicitly formulated before the beginning of the 20th century. Thus, the capability of human reason was significantly overvalued in the end of middle age and some scientific and ontological conclusions were taken as dogmas, which was criticized by other philosophers. On the other side, the scientific conclusion was being formulated also on the basis of other influences; e.g., F. Bacon (1561-1626) spoke about the idola, being based on some common convictions.  

During the new age the accession to science changed rather significantly also on the basis of different philosophical tendencies. It was influenced probably mainly by three philosophers: \\
 - R. Descartes (1596-1650) overvalued the role of human reason, which diminished the role of observation (including all characteristics of matter world); \\
 - D. Hume (1711-76) refused any causality and interpreted any evolution of matter world as based only on the chance; \\
 - A. Comte (1798-1857) refused any ontological metaphysics and limited all considerations to "positive" facts only, i.e. to mere measured values.
 
In the region of physics the positivist ideas were being propagated especially by E. Mach (1838-1916) who influenced the physics of the 20th century in decisive way. The physics of the 20th century was fully based on these ideas. 

\section {Physics of the 20th century}
The physics of the 20th century was represented mainly by special relativity theory and Copenhagen quantum mechanics. Both these theories have been based fully on phenomenological models. There is, however, a fundamental difference between them. While the causality is conserved in the relativity theory, the Copenhagen quantum mechanics is based on chance; and similar difference concerns also the interpretation and understanding of time. And one should ask and wonder in principle how it has been possible to apply both the theories to the same reality at the same time. It is evident that something should be wrong,

Even if it is possible to ask also how it is with the regularity of individual statements of the special relativity theory we will deal in the following with the problem of quantum mechanics; discussing  critical comments and arguments already published and analyzed in literature. And it is necessary to start with introducing at least main assumptions on which the Copenhagen quantum mechanics has been based:

 - first of all it is the time-dependent Schroedinger equation 
 \begin{equation}
  i\hbar\frac{\partial}{\partial t}\psi(x,t)=H\psi(x,t), \;\;\;\;
     H=-\frac{\hbar^2}{2m}\triangle + V(x)   \label{schr}
\end{equation}
where Hamiltonian $H$ represents the total (kinetic and potential) energy of a given physical system and $x$ represents coordinates of all matter objects; 
 
 - the evolution of a physical system is given by function $\psi(x,t)$ and all physical quantities may be determined as expected values of corresponding operators:
 \begin{equation}
           A(t) = \int\psi^*(x,t)A_{op}\psi(x,t)dx    
\end{equation}
where $A_{op}$ and functions $\psi(x,t)$ may be represented by operators and vectors of a suitable Hilbert space;

 - in the case of Copenhagen quantum mechanics it has been required for the corresponding Hilbert space to be spanned on one set of Hamiltonian eigenfunctions $\psi_E(x)$:
\begin{equation}
                H\psi_E(x) = E\psi_E(x);
\end{equation} 

 - and in addition to, the mathematical superposition principle valid in any Hilbert space has been interpreted in physical sense, i.e., any superposition of two physical states has represented again another basic (pure) physical state. 

It has been spoken often about two different interpretation alternatives of the Copenhagen mathematical model: orthodox (or Copenhagen) and statistical (or ensemble). However, it has been never mentioned that both the alternatives has corresponded to the different sets of assumptions. While Copenhagen alternative has involved all four introduced assumptions the ensemble alternative (denoted usually also as the hidden-variable theory) has corresponded to the first two assumptions only (being practically equivalent to the mere Schroedinger equation). And it is necessary to speak about two different theories.
 As to the Hilbert space in the latter alternative it must be chosen according to corresponding physical system; however, it must be always extended (at least doubled) in comparison to the Hilbert space required by the third assumption.

And we should ask, consequently, what is the actual difference between the classical physics and the Schroedinger equation (or the hidden-variable theory). 

\section {Schroedinger equation and classical physics}
E. Schroedinger \cite{sch} was successful with his equation when he showed that for particles exhibiting inertial motion the identical behavior with classical physics was obtained. 
Let us return now to the time-dependent Schroedinger equation
\begin{equation}
              i\hbar\frac{\partial}{\partial t}\psi(x,t)=H\psi(x,t)   \label{tsch}
\end{equation}
where the complex function $\psi(x,t)$ is expressed as
\begin{equation}
        \psi(x,t)  \;=\; \lambda(x,t)\, e^{\frac{i}{\hbar}\Phi(x,t)}  \label{psi}
\end{equation}
and both the functions $\lambda(x,t)$ and $\Phi(x,t)$ are real. Let us limit to time-independent potential $V(x)$.  Eq. (\ref{tsch}) may be substituted by two equations for two real functions (see D. Bohm \cite{bohm})  
\begin{eqnarray}
 \frac{(\nabla \Phi)^2}{2m}\,+\, V(x)\,+\,V_q(x,t)
                       &=& -\,\partial_t\,\Phi \;,   \label{hamj}   \\
 \triangle\Phi \,+\,2(\nabla\,\Phi)(\nabla\,lg\,\lambda) &=&
       -2m\;\partial_t\, lg\,\lambda   \label{ham2}
\end{eqnarray}
where
\begin{equation}
     V_q(x,t)\,=\,-\frac{\hbar^2}{2m}\frac{\triangle\lambda}{\lambda}
\end{equation}
has been denoted as quantum potential.   

Eq. (\ref{hamj}) resembles Hamilton-Jacobi equation 
\begin{equation}
      \frac{1}{2m}(\nabla S(x,t))^2 + V(x) = -{\partial_t S(x,t)}   \label{haja}
\end{equation}
where $S(x,t)$ has been replaced by $\Phi(x,t)$ and the quantum potential $V_q(x,t)$ has been added.  $S(x,t)$ is Hamilton principal function, from which the momentum values may be derived:
                \[   p(x,t)=\nabla S(x,t).  \]
For inertial motion it holds 
               \[ V_q(x,t)\,=\, V(x)\,=\,0  \]
and                               
              \[  \Phi(x,t)\,=\, S(x,t);  \]
the phase being identical with Hamilton principal function in such a case.              

Let us assume now
                       \[ V(x)\; \neq \; 0  \]
and let us limit to the basic solutions corresponding to different values of energy and  fulfilling the conditions
        \[  \psi_E(x,t)\,=\,\lambda_E(x)e^{-iEt},\;\;\; H\lambda_E(x)\,=\,E\lambda_E(x).  \]
In such a case $V_q(x)\neq 0$ is independent of $t$ and $\Phi(x,t)$ and $S(x,t)$ are mutually different; $V_q(x)$ representing the numerical measure of such difference.            
 There is not, however, any difference between the results of Schroedinger equation and classical physics; see \cite{adv}. All basic solutions of Schroedinger equation are equivalent to classical solutions corresponding to the same energy. However, it does not hold in opposite direction. In the case of discrete spectrum some solutions of Hamilton equations have not the corresponding counterparts in the Schroedinger equation.

In the past when the existence of quantum potential was related to physical difference there were done some attempts to interpret its existence as the consequence of Brown motions of individual microscopic objects. Our result is, however, in the full agreement with results of Ioanidou \cite{ioan} and Hoyer \cite{hoyer} who have shown that Schroedinger equation may be derived if classical physics is combined with a kind of statistical distribution. 
   
The advantage of the Schroedinger equation consists then in obtaining a complete statistical result in one solution if a statistical distribution of basic initial states is given. Consequently, Schroedinger equation is very suitable if some initial parameters (e.g., impact parameter in collision processes and, consequently, also total energy values) are not exactly known and only its statistical distribution may be established, as it occurs in measurement processes.  

\section {Quantum mechanics and critical comments}
Copenhagen quantum mechanics was accepted by physical community even if it exhibited some logical and ontological paradoxes and some critical comments were brought against its full regularity.
Already in 1933 Pauli \cite{pauli} showed that under the validity of all assumptions introduced in Sec. 4 it was necessary for the corresponding Hamiltonian to exhibit continuous energy spectrum from $-\infty$ to $+\infty$, which disagreed with the requirement of the energy being positive, or at least limited from below.
In 1964 Susskind and Glogover \cite{suss} showed then that exponential phase operator was not unitary, which indicated that the given Hilbert space was not complete to describe fully a corresponding physical system. 
Many unsuccessful attempts have been done to solve these deficiencies. The reason of the failure may be seen in the fact that practically in all cases both the shortages were taken and solved as one common problem.

The successful solution has been formulated only recently (see Refs. \cite{kund1,kund2}) when it has been shown that it is necessary to remove them one after the other. As to the system of two free colliding particles the Pauli problem may be removed if the Hilbert space required by the third assumption has been doubled as proposed by Lax and Phillips in 1967; it consists then of two mutually orthogonal subspaces 
(${\cal H}= {\it D}_{in} \oplus {\it D}_{out} $) being mutually linked with the help of evolution operator 
           \[  U(t) = e^{-iHt}.  \]
It holds
      \[ {\cal H} = {\overline {U(t)|{\it{D}}_{in} \rangle}} = {\overline {U(-t)|{\it{D}}_{out} \rangle}} \]
and similarly (see \cite{lax1,lax2}).
To remove also the problem of Susskind and Glogover the extended Hilbert space has had to be divided into two subspaces with opposite parities that have been connected with the help of exponential phase operator, with added action of its to bind mutually the zero energy states in the corresponding subspaces as proposed also already earlier by Fain \cite{fajn}; see also \cite{kund1,kund2}.    

The preceding problems concerned rather the mathematical structure of the Copenhagen model. Strong criticism based on physical features of the model was started by Einstein and his collaborators in 1935 \cite{epr}. They tried to show on the basis of a Gedankenexperiment that some non-physical (or rather anti-ontological) features were involved in Copenhagen quantum mechanics. They recommended to describe a microscopic object with the help of the so called hidden-parameter theory, which was refused, however, decisively by Bohr \cite{bohr35}. And the hidden-variable theory was refused also by physical community that accepted fully the phenomenological ideology.

The situation changed partially after the paper of Bohm \cite{bohm} and mainly the paper of J. Bell \cite{bell} were published. Bell derived some inequalities, the validity (or not) of which should have been tested experimentally and should have decided the controversy between Einstein and Bohr. Einstein's Gedankenexperiment was modified and corresponding experiments were realized; the coincidence probabilities of two photons transmitted through two polarizers lying in opposite directions were measured. The measurements were finished in 1982 with that the mentioned inequalities were violated \cite{asp}. The given result was interpreted as the refusal of hidden-variable theory and the victory of Bohr, which has lasted practically till now. 

However, we have succeeded in the last time to show that the given conclusion concerning the EPR experiments was done on the basis of two mistakes. First, it has not been taken into account that Bell's inequalities have been based on one important assumption that has limited strongly the range of their validity. In fact their violation in the given experiment has refused the classical physics only and not the hidden-variable theory. Second, it was also the mistake of Belinfante (see \cite{belin}) that influenced the given conclusion. Belinfante argued mistakenly that the predictions of Copenhagen quantum mechanics and hidden-variable theory in the given experiment  had to be necessarily different, which has not been true, either; see, e.g., \cite{abc}.

The mathematical model of microscopic phenomena described in Refs. \cite{kund1,kund2} (and corresponding to the hidden-variable theory) has removed, therefore, all discrepancies and has been in full harmony with Einstein's opinion.  \\

\section  {Other arguments against Copenhagen model}
The preceding arguments should be in principle sufficient to show that the hidden-variable theory (i.e., the Schroedinger equation without additional assumptions) should be strongly preferred to the Copenhagen quantum mechanics. However, two other important arguments may be added.

The first one concerns the fact that the further internal contradiction has been involved in the assumptions of Copenhagen model. It relates to the discrete part of Hamiltonian spectrum. On one side one argues that Schroedinger equation predicts the existence of states belonging to discrete energy values. And on the other side, the model requires the existence of pure states belonging to all intermediate energy values (i.e., to all superpositions). The experimental success concerning the Schroedinger equation was attributed in the past evidently to the whole Copenhagen model, even if two last assumptions were not involved in corresponding tests.

Now we should like to add another experimental argument refusing Copenhagen alternative. We shall start from the theory of polarized light, on which the EPR experiment has been based. And we shall measure the transmission of light through three polarizers, as indicated by the following scheme
                         \[       o---|---|^{\alpha}---|^{\beta}--->    \]
where $\alpha$  and  $\beta$ denote axis deviations of the second
and third polarizers from the first one. According to standard quantum-mechanical model it should hold for corresponding light transmission probability
     \[   P(\alpha,\beta)\,=\,\cos^2\alpha\,cos^2(\alpha-\beta). \]
The corresponding measurement was performed and the results were published in 1993-4; see \cite{krasa1,krasa2}. Fundamental deviations from the given quantum-mechanical formula have been found, as it may be seen from Fig. 1 and 2.

\begin{figure}[htb]
\begin{center}
\includegraphics*[scale=.32, angle=-90]{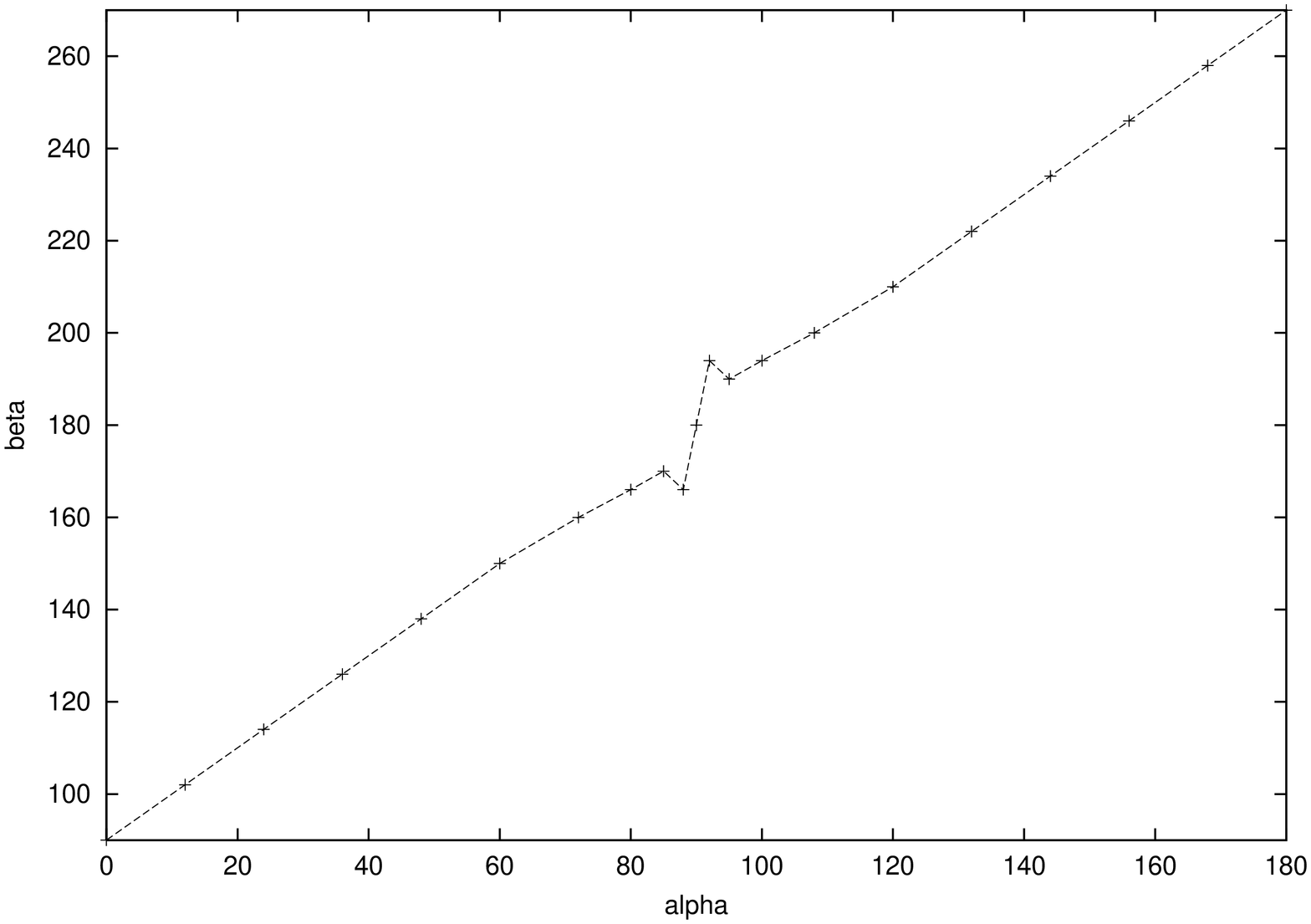}
\caption { \it {  The pairs of angles $\,\alpha$ and $\beta\,$ corresponding to minimal transmission values for chosen $\alpha$;
         pair values used for the measurement shown in Fig. 2.  } }
 \end{center}
% \end{figure}
% \begin{figure}[htb]
\begin{center}
\includegraphics*[scale=.32, angle=-90]{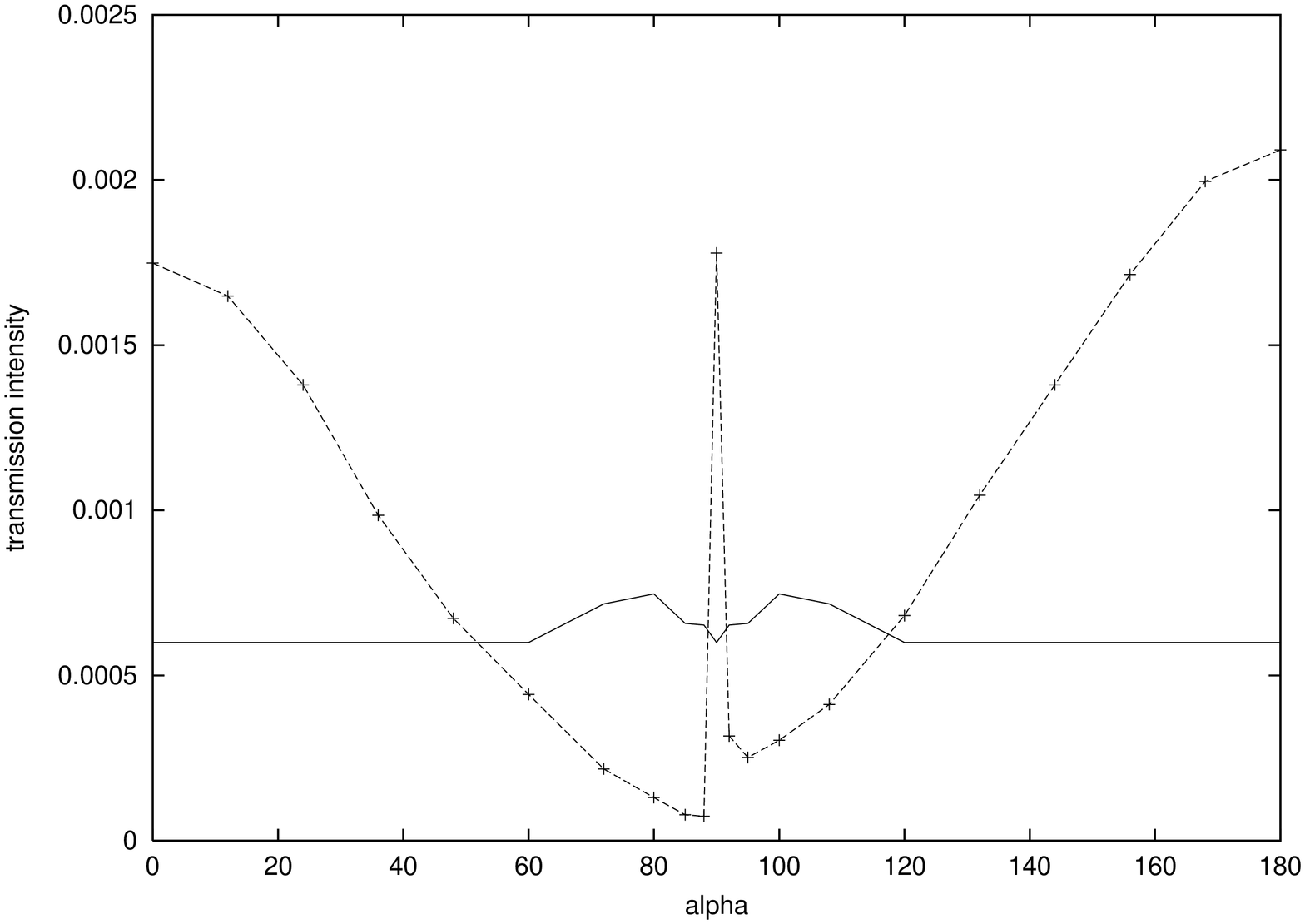}
  \caption { \it { Light transmission through three polarizers (for
     $\,\alpha\,-\,\beta\,$ pairs shown in Fig. 1); experimental data -
     points on dashed line; quantum-mechanical (orthodox) prediction -
     full line.     } }
 \end{center}
 \end{figure}
 
The experimental results (see \cite{krasa2}) obtained for angle combinations, shown in Fig. 1, are represented by dashed line in Fig. 2 (see also \cite{abc}). For a given angle $\alpha$ the angle $\beta$  was always established, so as the total light transmission be minimum. The quantum-mechanical prediction for these angle pairs is represented by full line; the position of this line being shifted in vertical
direction somewhat arbitrarily as the values of the so called "imperfectness" of given polarizers were not available. In any case the standard quantum-mechanical prediction requires maximums at the positions where the experiment exhibits deep minimums. The given discrepancy was not accented in \cite{krasa2} where the results were published for the first time as we were afraid reasonably that the paper would not have been accepted for publication in such a case.

It is evident that the Copenhagen quantum mechanics has represented the matter world very differently from the observed reality. R. Mirman has come recently to the same conclusion. In his book \cite{mirman} he has analyzed different conflicts of the Copenhagen picture with real world, which has led him to ask in the last section of this book whether the modern physics may be denoted yet as the science. 

As to the very recent attempts trying to bring a new support for Copenhagen alternative on the basis of recent delayed-choice experiment \cite {jac} we must stress that our preceding critical arguments have concerned the theoretical as well as experimental bases of the given alternative, which cannot be influenced by any other experimental data. The same holds for the attempts of A. Leggett \cite{legg} trying to extend the Copenhagen model to macroscopic area. In all such cases the hidden-variable theory fulfills all corresponding requirements. And the Copenhagen quantum-mechanical theory should be denoted as falsified.
 \\

\section  {Conclusion}
To conclude I should like to summarize shortly main points of the preceding text:

First, it is necessary to mention the method of human knowledge that consists in looking for falsification; there is not any verification. Scientific hypotheses or statements that have not been falsified may be denoted as plausible, which is to be related always to a kind of human belief.

Second, all analyzes of ours performed in connection with Copenhagen quantum mechanics have led clearly to preferring causality and ontological approach to mere phenomenological description of measured values. One is forced to return to scientific methods ruling from Aristotle till the end of the middle age.

And third, it is the preference of the hidden-variable theory (i.e., the proper Schroedinger equation) to Copenhagen quantum mechanics as well as to classical physics, even if the basic picture of classical matter world remains practically conserved.  \\

{\footnotesize

*) Extended version of the contribution to the Ninth International Symposium - Frontiers of Fundamental Physics, being held Jan. 7-9, 2008, University of Udine and ITCP Trieste (Italy) 

\end{document}